\newif\iftwocol
\newif\ifplacefig
\newif\ifdraft
\newif\ifcode

\twocoltrue
\placefigtrue
\draftfalse
\codetrue

\iftwocol
	\documentclass[apjl]{emulateapj}
\else
	\documentclass[preprint]{aastex}
\fi

\pdfoutput=1
\usepackage[normalem]{ulem}
\usepackage{color}
\usepackage{graphicx}
\usepackage{enumerate}
\ifdraft
	\usepackage{natbib}
	\bibpunct[,]{(}{)}{;}{a}{}{,}
\fi

\makeatletter

\newcommand{\vorovol}{TesseRACt}

\newcommand{\Msol}{\,M$_{\odot}$\,}

\newcommand{\myfig}[4]{%
\begin{figure}[ht]%
\begin{center}%
	\iftwocol%
		\ifdraft%
			\includegraphics[width=\columnwidth,keepaspectratio]{#2}%
		\else%
			\includegraphics[width=\columnwidth,keepaspectratio]{#1}%
		\fi%
	\else%
		\ifdraft%
			\includegraphics[width=\textwidth]{#2}%
		\else%
			\includegraphics[width=\textwidth]{#1.pdf}%
		\fi%
		\ifplacefig%
		\else%
			\vspace{-2.5in}%
		\fi%
	\fi%
	\caption{#3}%
	\label{#4}%
\end{center}%
\end{figure}%
}

\newcommand{\figconc}{%
\myfig{fig1}{../images/conc010}{%
Top: Concentration measured by \vorovol{ }and different spherical techniques as a function of known halo concentration. Bottom: \% relative residuals, 100*(measured-true)/true. The black line is the known concentration, blue is \vorovol, red is least-squares fitting to Eqn. \ref{eqn:nfw}, green is the half mass technique from Eqn. \ref{eqn:halfmass} assuming spherical symmetry, and magenta is the maximum circular velocity technique from \citet{Prada2012}. All of the techniques recover the correct value to within 2\%, regardless of concentration.%
}{fig:concen}
}

\newcommand{\figoblate}{%
\myfig{fig2}{../images/oblate010_c10}{%
Performance of \vorovol{ }vs. spherical techniques as a function of oblate halo ellipticity. While \vorovol{ }recovers concentration with 0.5\% of the correct value for all halo shapes, techniques assuming spherical symmetry do not and underestimate the concentration of the least spherical halos by up to 10\%.%
}{fig:oblate}
}

\newcommand{\figprolate}{%
\myfig{fig3}{../images/prolate010_c10}{%
Performance of \vorovol{ }vs. spherical techniques as a function of prolate halo ellipticity. \vorovol{ }recovers concentrations to within 0.3\% regardless of halo shape. The performance of techniques assuming spherical symmetry are highly dependent on halo shape and return less accurate halo concentrations by over 10\%. Overall, the spherical techniques are less accurate at recovering concentrations for prolate halos than oblate halos.%
}{fig:prolate}
}

\newcommand{\figtriax}{%
\myfig{fig4}{../images/triax010_c10_sqZ0p3}{%
Performance of \vorovol{ }vs. spherical techniques as a function of halo triaxiality. The dotted colored lines denote the fiducial value returned by each technique in the case where $T=1$ (an oblate halo with $e_{\rm oblate}=0.3$) and the residuals plotted here are with respect to this value. \vorovol{ }recovers concentrations within 0.2\% of the fiducial value for all triaxialities. The techniques which assume spherical symmetry display moderate dependence on triaxiality, but do not return concentrations that deviate from the fiducial value by more than 4\%.
}{fig:triax}
}

\newcommand{\figsubm}{%
\myfig{fig5}{../images/substr_mass010_c10_subr0p5_subc50_subrho0p5}{%
Performance of \vorovol{ }vs. spherical techniques as a function of subhalo mass. The concentrations returned by all techniques become less accurate as the mass of the subhalo increases. However, at the most massive end, \vorovol{ }is three times as accurate as techniques assuming spherical symmetry.%
}{fig:substr_mass}
}

\newcommand{\figsubr}{%
\myfig{fig6}{../images/substr_rsep010_c10_subm0p1_subc50_subrho0p5}{%
Performance of \vorovol{ }vs. spherical techniques as a function of subhalo location in the parent halo. All techniques perform poorly when the subhalo is close to the center of the parent halo. However, while \vorovol{ }becomes more accurate as the subhalo is placed further out, spherical techniques do not.%
}{fig:substr_rsep}
}

\newcommand{\figsubc}{%
\myfig{fig7}{../images/substr_conc010_c10_subm0p1_subr0p5_subrho0p5}{
Performance of \vorovol{ }vs. spherical techniques as a function of subhalo concentration. \vorovol{ }is more accurate than the spherical techniques overall, but exhibits some dependence on the concentration of the subhalo. The spherical techniques only exhibit a slight dependence for the subhalo that is less concentrated than the parent, where the non-parametric spherical techniques become more accurate.%
}{fig:substr_conc}
}

\newcommand{\fignpart}{%
\myfig{fig8}{../images/npart010_c10_sqY0p3_sqZ0p3}{%
Performance of \vorovol{ }vs. spherical techniques as a function of particle number. Dotted colored lines denote the fiducial value returned by each technique in the case of the full resolution, prolate halo. Residuals were computed with respect to this value. \vorovol{ }, fitting, and the half-mass technique are much less dependent on particle number than the technique which uses the peak velocity.%
}{fig:npart}
}

\begin{document}

\shorttitle{Non-parametric Halo Concentration}

\title{Voronoi Tessellation and Non-parametric Halo Concentration}

\author{Meagan Lang\altaffilmark{1}, Kelly Holley-Bockelmann\altaffilmark{1,\,2}, \& Manodeep Sinha\altaffilmark{1}}

\altaffiltext{1}{Department of Physics and Astronomy, Vanderbilt
University, Nashville, TN email: {\tt meagan.lang@vanderbilt.edu}} 

\altaffiltext{2}{Fisk University, Department
of Physics, Nashville, TN email:{\tt k.holley@vanderbilt.edu}}
%\altaffiltext{}{email: {\tt meagan.lang@vanderbilt.edu}}

%---------------------------------------------------------------------------------
% ABSTRACT
\begin{abstract}
We present and test \vorovol, a non-parametric technique for recovering the concentration of simulated dark matter halos using Voronoi tessellation. \vorovol{ }is tested on idealized N-body halos that are axisymmetric, triaxial, and contain substructure and compared to traditional least-squares fitting as well as two non-parametric techniques that assume spherical symmetry. \vorovol{ }recovers halo concentrations within 0.3\% of the true value regardless of whether the halo is spherical, axisymmetric, or triaxial. Traditional fitting and non-parametric techniques that assume spherical symmetry can return concentrations that are systematically off by as much as 10\% from the true value for non-spherical halos. \vorovol{ }also performs significantly better when there is substructure present outside $0.5R_{200}$. Given that cosmological halos are rarely spherical and often contain substructure, we discuss implications for studies of halo concentration in cosmological N-body simulations including how choice of technique for measuring concentration might bias scaling relations.
\end{abstract}

\keywords{galaxies: fundamental parameters --- galaxies: halos --- methods: numerical}
%\keywords{dark matter halos: concentration --- simulations: analysis techniques}

%---------------------------------------------------------------------------------
% INTRO
\section{Introduction}\label{S_intro}
There has long been a disconnect between the way we describe a dark matter halo and the reality of that dark matter structure. In theoretical terms, we think of a halo as a smooth, isotropic, virialized, usually spherical, typically uniformly rotating, distribution of mass that obeys a distinct radial distribution. However, while halos in cosmological simulations and observations may conform to these in a statistical sense, any one halo is not really any of these things. Halos are triaxial, anisotropic, and contain significant substructure \citep[e.g.][]{Jing2002,Gao2004,Despali2014,Groener2014}. Despite this well known fact, many analysis techniques try to extract halo properties by imposing one or more of these assumptions. For example, any procedure which fits a radial profile, be it Hernquist \citep{Hernquist1990}, NFW \citep{Navarro1996a}, or Einasto \citep{Einasto1989}, to a halos mass distribution assumes the halo is both spherical and smooth. In an era of simulations quickly surpassing $10^9$ particles, there is a need for physically motivated analysis techniques that do not impose constraints on what a halo should look like. 

Halo concentration is a particularly useful statistic for characterizing halos. Since halos that gain the majority of their mass at earlier times (when the mean density of the universe was higher) should be more compact, concentration is believed to encode a great deal of information about halo formation and growth. There have been numerous studies on the relationships between halo concentration in cosmological simulations and halo mass \citep{Navarro1996a,Neto2007}, redshift \citep{Bullock2001a,Gao2008,Klypin2011,Prada2012,Bhattacharya2013,Dutton2014,Ludlow2014,Diemer2015}, environment \citep{Bullock2001a,Maccio2007}, assembly history \citep{Wechsler2002,Zhao2009,Ludlow2013}, and cosmology \citep{Colin2000,Eke2001,Maccio2008,Dooley2014}. However, claims are often conflicting and the majority of techniques used to measure concentration fall victim to the above assumptions. 

We propose a non-parametric method for estimating halo concentration using Voronoi tessellation that we dub Tessellation based Recovery of Amorphous halo Concentrations (\vorovol). Section \S\ref{S_theory} briefly describes Voronoi tessellation and outlines \vorovol, \S\ref{S_tests} summarizes several tests, and \S\ref{S_discuss} summarizes our findings and describes studies that can benefit from \vorovol.

%---------------------------------------------------------------------------------
% METHODS
\section{Theory/Background}\label{S_theory}
%---------------------------------------------------------------------------------
% TRADITIONAL CONCENTRATION
\subsection{Measuring Concentration}
The concentration parameter is traditionally defined in an NFW halo as 
\begin{equation}
c_{\rm nfw} = \frac{R_{200}}{R_{\rm s}}
\end{equation}
where $R_{200}$ is the radius enclosing a mean density that is 200 times the critical density of the universe and $R_{\rm s}$ is the scale radius. Since $R_{200}$ can be easily found, concentration is typically obtained by fitting Eqn. \ref{eqn:nfw} to the radially enclosed mass profile to find $R_{\rm s}$.
\begin{eqnarray}
M_{\rm enc}(r) & = & 4\pi\rho_{0}R_{\rm s}^3\left[\ln\left(\frac{R_{\rm s}+r}{R_{\rm s}}\right)-\frac{r}{R_{\rm s}+r}\right]\label{eqn:nfw}
\end{eqnarray}

Although this is simple in theory, fitting even spherical halos without substructure can be difficult. Fits can be highly sensitive to resolution at the center, deviation from the expected NFW power laws, and choice of binning \citep[See][]{Prada2012}. These effects can be alleviated in practice by avoiding fitting altogether. Instead, the unknown profile parameters are related to other halo properties that can be robustly measured. For example, if $R_{\rm half}$ (the radius enclosing half the mass) and $R_{200}$ of a halo are known, it is possible to numerically solve  
\begin{equation}
\frac{1}{2} = \frac{\ln\left[(R_{\rm s}+R_{\rm half})/R_{\rm s}\right]-R_{\rm half}/(R_{\rm s}+R_{\rm half})}{\ln\left[(R_{\rm s}+R_{\rm 200})/R_{\rm s}\right]-R_{\rm 200}/(R_{\rm s}+R_{\rm 200})}\label{eqn:halfmass}
\end{equation}
for $R_{\rm s}$.

This can be done for any two independent halo properties, typically characteristic radii or velocities \citep{Avila-Reese1999,Thomas2001,Alam2002,Gao2004,Klypin2011,Prada2012}. While such techniques are more robust against deviations from NFW and yield more accurate results than fitting for dense halos with under sampled central regions, even these techniques still assume that halos are spherical and do not contain substructure. 

In principle, more accurate concentrations for non-spherical halos could be obtained by fitting to triaxial or ellipsoidal bins. Such techniques have been found to provide more accurate mass estimates in both simulations and observations, but generally assume that the axis ratio and alignment remains constant throughout the halo \citep{Warren1992,Jing2002,Allgood2006,Despali2013,Despali2014}. In reality, simulations show that halo shape is highly dependent on radius, becoming less and less spherical as you look deeper in the halo \citep{Allgood2006,Vera-Ciro2011}. This makes measurements assuming a constant shape dependent upon the location within the halo at which shape is measured. In addition, despite allowing for more freedom in halo shape, non-spherical fitting is still victim to the same caveats as spherical fitting and relies on the additional measurement of halo shape. 

While both non-spheroidal binning and non-parametric spherical techniques have advantages, neither is completely free of assumptions or can handle substructure.
However, using Voronoi tessellation, we can construct a technique that does not rely on fitting, does not make any assumptions of spherical symmetry, and allows for substructure.

%---------------------------------------------------------------------------------
% TESSELLATION CONCENTRATION
\subsection{Tessellation Based Concentration}
Given a set of seed points $\{p_1,\ldots,p_n\}$ in some space, Voronoi tessellation divides the space between the seeds such that each seed is the closest seed to its Voronoi region. In this way, each seed ($p_{i}$) has a corresponding Voronoi region of volume $V_{i}$ encompassing all points in space which are closest to that seed. Voronoi tessellation has been used to non-parametrically identify galaxy clusters in galaxy surveys \citep{Soares-Santos2011}, identify dark matter halos \citep{Neyrinck2004} and voids \citep{Neyrinck2008} in cosmological simulations, and improve the treatment of hydrodynamics in simulations \citep{Mocz2013,Hopkins2014}. We take this one step further. Once halos are identified in cosmological simulations (either by friends-of-friends, spherical over-density, or tessellation), the additional information provided by the particles' associated volumes can be used to derive halo properties (like concentration) without imposing any additional functional form.

To determine concentration from the particle volumes, a profile is constructed that describes how mass scales with volume rather than radius. For a particle $p_{i}$ with mass $m_{i}$ and a Voronoi volume $V_{i}$, the volume $V_{{\rm enc},i}$ `enclosed' by that particle is taken to be the sum of all particle volumes which are smaller than $V_{i}$ or
\begin{equation}
V_{{\rm enc},i} = \sum_{j=0}^{n}V_{j}[V_{j}\leq V_{i}].
\end{equation}
Similarly, the mass $M_{{\rm enc},i}$ `enclosed' by a particle $p_{i}$ is  
\begin{equation}
M_{{\rm enc},i} = \sum_{j=0}^{n}m_{j}[V_{j}\leq V_{i}].
\end{equation}
Each particle can then be assigned a theoretical `radius' $R_{i}'=(3V_{{\rm enc},i}/4\pi)^{1/3}$ that is defined as the radius the particle would be at if $V_{{\rm enc},i}$ were spherical. The result is a volume based mass profile $M_{\rm enc}(R')$. Naively, the volume based concentration could then be defined as 
\begin{equation}
c_{\rm vol} = \left(\frac{V_{200}}{V_{\rm s}}\right)^{1/3}
\end{equation}
where $V_{200}$ is the densest volume containing an average density that is 200 times the critical density of the universe and $V_{\rm s}$ is some scale volume. If the theoretical radii ($R_{200}'$ and $R_{\rm s}'$) associated with these volumes converged to the corresponding physical radii ($R_{200}$ and $R_{\rm s}$) in the case of a spherical halo, $c_{\rm vol}$ would equal $c_{\rm nfw}$ and this would be sufficient. However, this is not strictly true.

For even a spherical halo, the relationship between a particle's physical radius $R_{i}$ and theoretical radius $R_{i}'$ is not 1:1. Due to the intrinsic scatter in the inter-particle spacing at a given physical radius, particles with slightly larger/smaller volumes will be scattered to larger/smaller theoretical radii. Because the density of particles is always greater toward smaller physical radii, it is more likely for particles inside $R_{i}$ to be scattered to larger theoretical radii. As a result, there will then be systematically fewer particles considered `enclosed' by particle $p_{i}$ and $R_{i}'$ will be systematically lower.

In order to correct for this and preserve the same numerical values for $c_{\rm vol}$ and $c_{\rm nfw}$ in the case of spherical halos, the volume based concentration is defined as
\begin{eqnarray}
c_{\rm vol} & = & \beta\left(\frac{V_{\rm 200}}{V_{\rm s}}\right)^{\alpha/3}. \label{eqn:cvol0}
\end{eqnarray}
$\beta=0.8062$ and $\alpha=1.0417$ were obtained by fitting to measurements of $V_{\rm 200}$ and $V_{\rm s}$ for 10 spherical halos with known concentrations between 5 and 70. While this treatment is simplistic, tests performed in \S\ref{SS:conc} indicate that it should be sufficient for most studies.

Eqn. \ref{eqn:cvol0} can then be simplified in terms of the radii a halo would have if it were spherical. 
\begin{eqnarray}
c_{\rm vol} & = & \beta\left(\frac{R_{\rm 200}'}{R_{\rm s}'}\right)^{\alpha}. \label{eqn:cvol}
\end{eqnarray}
Note that since $V_{\rm 200}$ is defined as the densest volume containing an average density that is 200 times the critical density of the universe, $R_{\rm 200}'$ can be found directly from the Voronoi volumes. However, $V_{\rm s}$ and $R_{\rm s}'$ are somewhat arbitrary in the absence of fitting. Instead, we choose to define the scale radius and volume in terms of a quantity we can measure. We adopt $V_{\rm half}=4\pi R_{\rm half}'^3/3$, the densest volume enclosing half the mass. 

Once $R_{\rm half}'$ is known, the corresponding $R_{\rm s}'$ for a spherical halo can be found by numerically solving Eqn. \ref{eqn:halfmass}. This relationship yields unique concentrations for each $R_{\rm half}$ when $1<c<100$. Above $c=100$, $R_{\rm half}/R_{\rm 200}$ becomes somewhat degenerate about 0.1.

The \vorovol{ }procedure for finding $c_{\rm vol}$ is then:
\begin{enumerate}
	\setlength\itemsep{0cm}
	\item Run Voronoi tessellation to determine the volumes and densities at each particle.
	\item Rank particles in order of decreasing density (increasing volume).
	\item Calculate enclosed mass ($M_{\rm enc}$) and volume ($V_{\rm enc}$) at each particle by summing the masses and volumes of `denser' particles.
	\item Calculate mean enclosed density at each particle ($\left<\rho_{\rm enc}\right>=M_{\rm enc}/V_{\rm enc}$)
	\item Find $V_{\rm 200}$ and $M_{\rm 200}$ (the volume and mass enclosed by the particle at which the mean enclosed density reaches 200 times the critical density of the universe)
	\item Find $V_{\rm half}$ (the volume enclosed by the particle at which the enclosed mass reaches $M_{\rm 200}/2$)
	\item Calculate $R_{\rm 200}'$ and $R_{\rm half}'$ from $V_{\rm 200}$ and $V_{\rm half}$
	\item Numerically solve Eqn. \ref{eqn:halfmass} for $R_{\rm s}'$
	\item Calculate $c_{\rm vol}$ from Eqn. \ref{eqn:cvol}
\end{enumerate}

%---------------------------------------------------------------------------------
% TESTS
\section{tests}\label{S_tests}
%To compare the ability of \vorovol{ }to recover concentrations against traditional fitting and two non-parametric techniques that assume spherical symmetry, 
%
For the tests that follow, we compare our volume-based non-parametric technique \vorovol{ }to both least-squares NFW fitting, as well as two non-parametric techniques that assume spherical symmetry. The first infers the scale radius from the $R_{\rm half}$ and $R_{200}$ in the same fashion as Eqn. \ref{eqn:halfmass}, but using the physical particle radii. The second uses the relationship between the peak circular velocity and circular velocity at $R_{200}$ from \citet{Prada2012},
\begin{equation}
\frac{v_{\rm max}}{v_{\rm 200}} = \left[\frac{0.216c}{\ln(1+c)-c/(1+c)}\right]^{1/2}.
\end{equation}

For each test, we run all four techniques on isolated N-body halos generated by sampling spherical NFW profiles of known concentration. Unless otherwise stated, each halo has $N_{\rm part}=1\times10^6$ particles, $M_{\rm 200}=1\times10^{12}$\Msol, and $R_{\rm 200}=200$\,kpc. The halos only differ in $\rho_{0}$ and $R_{\rm s}$.

To ensure that insufficient sampling of the profile does not affect the accuracy of the least squares fitting, fits are bounded on the lower end at $0.05R_{\rm 200}$ and $R_{\rm 200}$ on the upper end.

Errors on concentration measurements were determined by running each technique on 10 different realizations of each test halo generated by using a different random number seed. In the tests below, the mean concentration returned by each technique is plotted with the standard deviation across the different realizations.

%---------------------------------------------------------------------------------
% CONCENTRATION TEST
\subsection{Concentration}\label{SS:conc}
We first tested the performance of each method on halos with concentrations of 5, 10, 25, and 50. Results are shown in Figure \ref{fig:concen}.
%
% Figure 1 Here
\ifplacefig
	\figconc
\fi

As these are idealized halos, all of the tested techniques return accurate measures of concentration as expected. For the idealized halos, \vorovol{ }is the most accurate for all but the $c=5$ halo for which the measured concentration is still within 2\% of the correct value. Of the techniques which assume spherical symmetry, all three perform similarly with the maximum circular velocity and fitting techniques returning marginally more accurate concentrations at the low and high ends respectively. The accuracy of the techniques which assume spherical symmetry does not appear to be overly concentration dependent. However, as the modified concentration definition used for \vorovol{ }in Eqn. \ref{eqn:cvol} was fit to measurements across a range of concentrations, \vorovol{ }is slightly more accurate for halos with intermediate concentrations. All of the techniques tested were consistent across the 10 halo realizations with standard deviations of $<0.6$\%.

%---------------------------------------------------------------------------------
% HALO SHAPE TESTS
\subsection{Halo Shape}\label{SS_shapetests}
%---------------------------------------------------------------------------------
% OBLATE HALO TESTS
\subsubsection{Oblateness}
To test how each method performed on oblate halos, the $c=10$ halo from above was squeezed along the $z$ axis using the volume preserving transformation
\begin{eqnarray}
\left[\begin{array}{c}x'\\y'\\z'\end{array}\right] & = & \left(\frac{a}{b}\right)^{1/3}
\left[\begin{array}{ccc}
a & 0 & 0 \\
0 & a & 0 \\
0 & 0 & b 
\end{array}\right]
\left[\begin{array}{c}x\\y\\z\end{array}\right],
\end{eqnarray}
for 7 different values of ellipticity where $e_{\rm oblate}=b/a$ and $a=1$. Results are shown in Figure \ref{fig:oblate}.
%
% Figure 2 Here
\ifplacefig
	\figoblate
\fi

For all but the least spherical halos, \vorovol{ }consistently recovers concentrations within 0.5\% of the correct value, with a minor dependence on halo shape (overestimated by 0.4\% for $e_{\rm oblate}=0.3$ vs. underestimate by 0.02\% for spherical halos). However, the performance of the techniques which assume spherical symmetry is highly dependent on halo shape. Fitting performs the worst overall, with increasingly underestimated values for the least spherical halos (10\% for $e_{\rm oblate}=0.3$). The performance of the two non-parametric spherical techniques is dependent on halo shape in a more complicated manner. 

For decreasing values of $e_{\rm oblate}$ (decreasing spherical symmetry), both non-parametric spherical techniques have moderately increasing positive residuals until around $e_{\rm oblate}=0.6$ to 0.7 where the residuals begin to rapidly decrease resulting in underestimates of 6 and 8\% for the half-mass and peak-velocity techniques respectively. These dependencies arise because the deformation of the halo flattens the radial mass profile. As fitting uses the entire profile, the concentration is underestimated for all non-spherical halos. However, non-parametric techniques use only two points in the profile ($R_{200}$ and some inner radius) and their performance will depend up how these parts of the profile are affected by the transformation. $R_{200}$ continually increases as compression of the halo edge decreases the density at the edge of the profile. The inner radius will increase as well, but at a slower rate resulting in an overestimation of the concentration until the edge of the halo is compressed to a size comparable to the inner radius. Once this occurs, the inner radius increases more rapidly than $R_{200}$ resulting in underestimates. The precise value of $e_{\rm oblate}$ at which this occurs will depend on the radius that a particular non-parametric technique utilizes. Since the velocity peaks at a smaller radius than the half-mass radius for an NFW profile, the technique which uses the peak velocity is less sensitive to this effect.

The test halos here have the same ellipticity at all radii by design. However, real halos are found to be decreasingly spherical at their centers \citep{Allgood2006,Vera-Ciro2011}. While \vorovol{ }would not be affected, it is likely that the spherical techniques would produce less accurate concentrations. This would be particularly pronounced for fitting due to sensitivity to the inner profile slope.

%---------------------------------------------------------------------------------
% PROLATE HALO TESTS
\subsubsection{Prolateness}

To test how each method performed on prolate halos, the $c=10$ halo from above was squeezed along it's $z$ and $y$ axes using the volume preserving transformation
\begin{eqnarray}
\left[\begin{array}{c}x'\\y'\\z'\end{array}\right] & = & \left(\frac{a^2}{b^2}\right)^{1/3}
\left[\begin{array}{ccc}
a & 0 & 0 \\
0 & b & 0 \\
0 & 0 & b 
\end{array}\right]
\left[\begin{array}{c}x\\y\\z\end{array}\right],
\end{eqnarray}
for 7 different values of ellipticity where $e_{\rm prolate}=b/a=c/a$ and $a=1$. Results are shown in Figure \ref{fig:prolate}.
%
% Figure 2 Here
\ifplacefig
	\figprolate
\fi

The performance across all techniques is similar to the above tests for oblate halos. However, there does appear to be a stronger dependence on ellipticity at intermediate values for prolate halos than for oblate halos for techniques which assume spherical symmetry. This is because the prolate transformation flattens the radial mass profile to a greater degree than the oblate transformation. 

%---------------------------------------------------------------------------------
% TRIAXIAL HALO TESTS
\subsubsection{Triaxiality}

We then tested the performance of each method on triaxial halos (See Figure \ref{fig:triax}). For 9 different values of triaxiallity \citep[$T=\frac{c^2-b^2}{c^2-a^2}$, for $c\leq b\leq a$,][]{Franx1991}, the $c=10$ halo from above was squeezed along the $z$ and $y$ axes using the volume preserving transformation
\begin{eqnarray}
\left[\begin{array}{c}x'\\y'\\z'\end{array}\right] & = & \left(\frac{a^2}{bc}\right)^{1/3}
\left[\begin{array}{ccc}
a & 0 & 0 \\
0 & b & 0 \\
0 & 0 & c 
\end{array}\right]
\left[\begin{array}{c}x\\y\\z\end{array}\right],
\end{eqnarray}
where $a=1$, $c/a=0.3$, and $b=\sqrt{c^2-T(c^2-a^2)}$.
%
% Figure 4 Here
\ifplacefig
	\figtriax
\fi

\vorovol{ }out performs all spherical techniques at $<0.2$\% for all triaxialities. The spherical techniques return concentrations that are only slightly less accurate overall ($<$4\%) than in the case where $T=1$. While \vorovol's performance does not depend on triaxiality, the accuracy of the techniques which assume spherical symmetry does to a small degree. The two non-parametric techniques become less accurate for lower values of $T$, while fitting only exhibits this behavior until $T\sim0.3$, where this trend is reversed. However, since the residuals never exceed 4\%, it is reasonable to conclude that only the ratio between the largest and smallest halo axes appears to have a significant impact on the accuracy of the concentration estimate.

%---------------------------------------------------------------------------------
% SUBSTRUCTURE TESTS
\subsection{Halos with Substructure}
To test the ability of \vorovol{ }to recover concentrations in the presence of substructure, idealized subhalos of varying mass and concentration were added to the $c=10$ halo at varying radii. In each case the mass of the subhalo was set by downsampling the test halo from the previous section with the desired concentration. The size of the subhalo was then scaled such that the mean density within $0.2R_{\rm 200}$ was half that of the parent halo.

%---------------------------------------------------------------------------------
% SUBSTRUCTURE MASS TESTS
\subsubsection{Substructure Mass}
Figure \ref{fig:substr_mass} shows the results from varying the mass of the subhalo across 1-20\% of the parent halo's $M_{200}$. The subhalos had a concentration of 50 and were placed $0.5R_{\rm 200}$ from the center of the parent halo. Since subhalos were scaled to have a constant central density, increasing the mass of the subhalo also increases its size. 
% Figure 5 Here
\ifplacefig
	\figsubm
\fi

As expected, all techniques become less accurate as the subhalo increases in mass and size, but \vorovol{ }performs significantly better. For the most massive subhalo, 20\% the mass of the parent, \vorovol{ }overestimates the concentration by 8\%, while techniques assuming spherical symmetry underestimate concentrations by up to 30\% of the true value. \vorovol{ }overestimates concentrations because particles in the subhalo have small enough volumes that they are assumed to be within the inner parts of the parent halo and thus result in a smaller half-mass volume. The techniques assuming spherical symmetry underestimate concentrations in the presence of substructure because the subhalo contributes mass to the outer profile.

%---------------------------------------------------------------------------------
% SUBSTRUCTURE RADIUS TESTS
\subsubsection{Substructure Radius}
Figure \ref{fig:substr_rsep} shows the results from varying the distance of the subhalo from the center of the parent halo from 0.05 to 0.5$R_{\rm 200}$. The subhalos each had a concentration of 50 and 1\% the mass of the parent halo. 
%
% Figure 6 Here
\ifplacefig
	\figsubr
\fi

At small radii, the subhalo almost coincides with the center of the parent halo, causing all techniques to return higher concentrations than they would had there been no substructure present. As the subhalo is placed at larger radii, the concentrations returned by all techniques become lower. However, while concentrations returned by techniques assuming spherical symmetry continue to drop past the true value as additional mass is added to the outer profile resulting in underestimates by up to 20\%, \vorovol{ }becomes more accurate, underestimating the concentration by only 0.7\% at 0.75$R_{\rm 200}$. This occurs because, when the subhalo is placed at larger radii where the density of the parent is lower, the densities assigned to the subhalo particles using tessellation become lower and contribute less to the half-mass volume. The radius at which the spherical techniques change from overestimating to underestimating depends upon which part of the mass profile is used to calculate concentration.

%---------------------------------------------------------------------------------
% SUBSTRUCTURE CONCENTRATION TESTS
\subsubsection{Substructure Concentration}
Figure \ref{fig:substr_conc} shows the results from varying the concentration of the subhalo from 5 to 50. The subhalos were placed $0.5R_{\rm 200}$ from the center of the parent halo and had 1\% the mass of the parent. 
%
% Figure 7 Here
\ifplacefig
	\figsubc
\fi

For subhalos of equal or greater concentration than the parent ($c_{\rm sub}\geq10$), \vorovol{ }is more accurate than the spherical techniques ($<3$\% vs. $<$20\%). While the accuracy of the spherical techniques shows little dependence on concentration in this regime, \vorovol's accuracy changes with concentration. Since \vorovol{ }places no constraints on where the center of the halo is, the measured concentration is essentially an average of the concentrations of the two halos present, the parent and the subhalo. As a result, the same dependence on halo concentration that was seen in \S\ref{SS:conc} is also present for subhalo concentration.

The case of a subhalo that is less concentrated than the parent halo ($c_{\rm sub}=5$) is less physical. In order to maintain a central density half that of the $c=10$ parent halo with 1\% of the mass, the $c=5$ halo must be scaled to less than half of its original size. This means that the substructure contributes to a narrower region of the radial mass profile, resulting in lower estimates for $R_{200}$ and $M_{200}$ that are closer to those for the parent halo alone. For fitting, which uses the entire profile, the result is a significantly underestimated concentration (20\%). However, the non-parametric techniques, which are not dependent on the whole profile, are significantly more accurate in this case. The half-mass technique is slightly more accurate (2\%) than the peak velocity technique (3\%) because while $R_{\rm half}$ changes with $R_{200}$, the peak velocity occurs well within the radius at which the substructure was placed. 

%---------------------------------------------------------------------------------
% RESOLUTION TESTS
\subsection{Dependence on Particle Number}
Finally, we tested the performance of each technique for different resolutions (See Figure \ref{fig:npart}). Beginning with the $c=10$ halo containing $10^6$ particles, transformed to be prolate with an $e_{\rm prolate}=0.3$, random subsets of the particles were selected to create halos of the same size and shape, but lower resolution. 
%
% Figure 8 Here
\ifplacefig
	\fignpart
\fi

Above $2\times10^3$ particles, all of the tested techniques are reasonably convergent ($<$10\%) around the fiducial concentration for $N=1\times10^{6}$. For fewer particles, the residuals for the peak velocity technique climb quickly to $>80$\% of the fiducial value for 100 particles. \vorovol{ }and the half-mass non-parametric technique perform similarly on average, returning concentrations within 20\% of the fiducial value at the lowest resolution. However, \vorovol{ }was more precise than the half-mass technique. Across the 10 halo realizations, the standard deviation in the concentrations returned by \vorovol{ }and the half-mass technique were 40\% and 50\% respectively at $N=100$. Fitting was surprisingly accurate at low resolutions with residuals $<10$\% and standard deviations similar to \vorovol.

%---------------------------------------------------------------------------------
% SUMMARY & DISCUSSION
\section{Summary \& Discussion}\label{S_discuss}

For idealized spherical halos, \vorovol{ }is slightly more accurate at recovering intermediate concentrations for N-body halos than techniques that assume spherical symmetry ($\sim$0.5\% vs. $\sim$1\%). However, \vorovol{ }truly shines for non-spherical halos. For the most oblate or triaxial galaxies, even the most accurate spherical technique using the peak circular velocity returned concentrations that underestimated the true value by up to $\sim$10\%, while \vorovol{ }had residuals of only $\sim$0.5\%. This is troubling given that halos in simulations are not often spherical. 

Studies of halos in cosmological simulations indicate that halos tend to be prolate overall, halos become increasingly triaxial at higher redshifts, and the most massive halos at all redshifts are the least spherical \citep{Allgood2006,Bett2007,Vera-Ciro2011,Despali2014}. Therefore, concentrations resulting from techniques that assume spherical symmetry would result in median concentrations that are biased overall to be lower than the true value and especially biased for halos that are massive or at high redshift. This bias could be expressed in the scaling relations between concentration and other intrinsic halo properties like mass, formation redshift, and environment.

The concentration of a halo is believed to reflect the state of the universe at its collapse \citep{Navarro1996a}. Therefore, on average, more massive halos should have lower concentrations, halos at higher redshift should have lower concentrations than those of the same mass at lower redshift, and halos in denser environments should have higher concentrations. However, the exact relationship between halo concentration, mass, redshift, and environment is still under debate \citep{Bullock2001a,Eke2001,Neto2007,Gao2008,Klypin2011,Prada2012,Bhattacharya2013,Dutton2014,Ludlow2014,Diemer2015}. Since halo shape also varies systematically with mass and redshift, it is possible that imposing the assumption of spherical symmetry could systematically bias such studies. Use of a non-parametric technique that is independent of halo shape like \vorovol{ }could be instrumental in eliminating this bias and pinning down these relations.

In addition to biasing the median measured concentration for halos in a given mass and redshift bin, the error introduced by assuming spherical symmetry would also increase scatter in the distribution of concentration for these halos due to scatter in halo shape. Intrinsic scatter in this relationship is expected to result from differences between any two halos' mass assembly histories that result in slightly different concentrations. However, if scatter is also being introduced by the technique used to measure concentration, measurement of this scatter becomes less informative. Since scatter in the concentration-mass relationship is greatest for those halos which are the least virialized and have the poorest NFW fits \citep{Jing2000,Neto2007}, removing these halos can help to remove some of this bias. However, even virialized halos can be non-spherical and it is likely that the assumption of spherical symmetry would still contribute to the remaining scatter. \vorovol{ }could help to identify what the true intrinsic scatter in the concentration-mass relationship is for both virialized and un-virialized halos.

\vorovol{ }could be improved for future studies in a number of ways. One possibility is using tessellation in the full six dimensional phase space to place better constraints on how deep a particle resides within the host galaxy's potential. This would be particularly useful for correcting for substructure. While substructure particles may be confused with parent halo particles on the basis of spatial volumes alone, the two would be more easily differentiated in velocity space. It may also be possible to improve \vorovol's performance by more accurately parameterizing the relationship between traditional concentration and that calculated using tessellation based volumes as in Eqn. \ref{eqn:cvol}. Fitting over a wider range of concentrations and allowing a more complex relationship should improve the slight dependence of \vorovol's performance on concentration. However, since this dependence only affects concentrations by $<3$\%, the parameterization presented here should be sufficient for most studies.

Although only spatial information was used here, tessellation can also be used to glean information non-parametrically from simulations by tessellating over other parameters. In simulations which include gas, tessellation over parameters like star formation rate, temperature, and metallicity in addition to position and velocity can help to identify structure and explore formation scenarios. 

%---------------------------------------------------------------------------------
% CODE COMMENT FOR ARXIV
\ifcode%
\section{Code}\label{S_code}
The \vorovol{ }method for estimating concentrations from Voronoi volumes has been assembled into a publicly available Python package that will be made available at the time of publication. This release will include example halos, routines for running and handling output from an adapted version of VOBOZ \citep{Neyrinck2004}, and routines for estimating concentration using the spherical techniques tested here. %
\fi%

%---------------------------------------------------------------------------------
% ACKNOWLEDGMENTS
\acknowledgments
K.H-B. acknowledges the support of a National Science Foundation Career Grant
AST-0847696 as well as the supercomputing support of Vanderbilt's Advanced Center for Computation Research and Education. Support for M.L. was provided by a National Science Foundation Graduate Research Fellowship.

%---------------------------------------------------------------------------------
% REFERENCES
\ifdraft
	\bibliography{Flybys.bib}
\else

\fi

%---------------------------------------------------------------------------------
% FIGURES
\ifplacefig
\else
	\figconc
	\figoblate
	\figprolate
	\figtriax
	\figsubm
	\figsubr
	\figsubc
	\fignpart
\fi

%---------------------------------------------------------------------------------
% COMMENTS
%\newpage
%\todo{General things to do:
%\begin{itemize}
%\end{itemize}
%}

\end{document}